# Measuring Nonlinear Relationships and Spatial Heterogeneity of Influencing Factors on Traffic Crash Density Using GeoXAI

**Jiaqing Lu**

Department of Civil and Environmental Engineering

Florida State University

2525 Pottsdamer St, Tallahassee, FL 32310 Email: jl23br@fsu.edu

**Ziqi Li**

Department of Geography

Spatial Data Science Center

Florida State University

113 Collegiate Loop, Tallahassee, FL 32306 Email: ziqi.li@fsu.edu

**Lei Han**

Department of Civil, Environmental & Construction Engineering

University of Central Florida

Orlando, FL 32816, Email: le966091@ucf.edu

**Qianwen Guo\***

Department of Civil and Environmental Engineering

Florida State University

2525 Pottsdamer St, Tallahassee, FL 32310 Email: qguo@fsu.edu

# Abstract

This study applies a Geospatial Explainable AI (GeoXAI) framework to analyze the spatially heterogeneous and nonlinear determinants of traffic crash density in Florida. By combining a high-performing machine learning model with GeoShapley, the framework provides interpretable, tract-level insights into how roadway characteristics and socioeconomic factors contribute to crash risk. Specifically, results show that variables such as road density, intersection density, neighborhood compactness, and educational attainment exhibit complex nonlinear relationships with crashes. Extremely dense urban areas, such as Miami, show sharply elevated crash risk due to intensified pedestrian activities and roadway complexity. The GeoShapley approach also captures strong spatial heterogeneity in the influence of these factors. Major metropolitan areas including Miami, Orlando, Tampa, and Jacksonville display significantly higher intrinsic crash contributions, while rural tracts generally have lower baseline risk. Each factor exhibits pronounced spatial variation across the state. Based on these findings, the study proposes targeted, geography-sensitive policy recommendations, including traffic calming in compact neighborhoods, adaptive intersection design, speed management on high-volume corridors such as I-95 in Miami, and equity-focused safety interventions in disadvantaged rural areas of central and northern Florida. Moreover, this paper compares the results obtained from GeoShapley framework against other established methods (e.g., SHAP and MGWR), demonstrating its powerful ability to explain nonlinearity and spatial heterogeneity simultaneously.

# 1. Introduction

The persistent toll of traffic crashes on public health and economic vitality remains a significant challenge for modern societies (WHO, 2022). In the United States, the economic costs of crashes averaged $340 billion over a five-year period, including lost productivity, medical care, and property damage (Blincoe et al., 2023). This financial toll is coupled with a tragic human cost, with over 40,000 people killed on U.S. roadways in 2023 (National Center for Statistics and Analysis, 2025). These numbers highlight the urgent need for analytical approaches that can address the complex factors that contribute to crashes. Gaining a deeper understanding of what drives crash risk is therefore crucial for guiding more effective safety policies.

Socioeconomic, traffic operations, and roadway infrastructure characteristics have been found to strongly influence crash patterns across regions. For example, communities with higher proportions of economically disadvantaged or age-extreme populations often face elevated crash risk due to factors such as limited access to safe vehicles, driver training, or pedestrian facilities, as well as inexperience or mobility impairments (Males, 2009; Huang et al., 2010; Sohaee and Bohluli, 2024). Traffic volume, truck presence, roadway geometry, and speed-related factors further affect crash likelihood and severity, while enforcement mechanisms like speed cameras and traffic signals can mitigate risks (Dong et al., 2015; Zou et al., 2017; Rahman et al., 2018; De Pau et al., 2014; Polders et al., 2015; Goswamy et al., 2023). Together, these socio-economic and traffic characteristics shape the spatial and temporal distribution of traffic crashes.

How these factors influencing crash risk can be complicated. There are two main effects that are being reported in the crash modelling literature: non-linearity and spatial heterogeneity. Crash occurrences are often governed by complex, nonlinear relationships with factors such as traffic volume and road geometry. For example, traffic volume may initially reduce crash frequency up to a certain threshold due to congestion but increase it beyond that point due to exposure. These threshold effects between variable interactions challenge traditional linear modeling assumptions. Classical regression methods such as Poisson or Negative Binomial regression cannot capture such complex nonlinear relationships between variables (e.g., traffic volume) and crash frequency. While polynomial transformations, piecewise regression, and Generalized Additive Models (GAMs) can partially address nonlinearity, they often struggle with high-dimensional or highly nonlinear interactions (Park & Lord, 2007; Zhang et al., 2012). In contrast, Machine learning (ML) models such as random forests, gradient boosting, and neural networks excel at modeling complex, nonlinear, and high-dimensional relationships in crash analysis (An et al., 2022; Cui et al., 2025; Han & Abdel-Aty, 2025). With the advances in Explainable AI such as SHAP allow interpretable insights from these black-box models by quantifying feature importance and revealing the functional forms of relationships (Lundberg & Lee, 2017). For instance, Zhang et al. (2023) integrate ML and SHAP to identify how weather, lighting, and road geometry influenced crash severity. Similarly, Xue et al. (2024) apply SHAP-based analysis to quantify the contribution of vision-based features from Google Street View to cycling crash risk at intersections in Manhattan.

Regarding spatial heterogeneity, the influence of factors, such as roadway or demographic characteristics, is rarely uniform across all places (Liu et al., 2017; Wang et al., 2024). Crash occurrences and their contributing factors often exhibit strong spatial heterogeneity due to variations in the built environment, driver behavior, and socio-demographic conditions across locations. For example, a dense road network may impact crash density very differently in a compact urban area compared to a sprawling suburban area, a phenomenon known as spatial heterogeneity (Xiao et al., 2024). Analyzing and modeling this spatial heterogeneity is crucial for uncovering localized trends and understanding how various factors interact across geographic areas. Spatial heterogeneity is often modelled by spatial statistical models such as Geographically Weighted Regression (GWR) models and mixed models with spatial random effects (Liu et al, 2017; Tang et al., 2020).

Even though both non-linearity and spatial heterogeneity are recognized and reported in the crash literature, there is a lack of methods and studies that can simultaneously address both aspects. This is primarily because methods that capture non-linearity often overlook spatial heterogeneity, while spatial statistical models that explicitly measure spatial heterogeneity generally do not specify non-linearity, which is otherwise difficult to estimate. Consequently, this study addresses this critical gap by simultaneously identifying spatial heterogenous and non-linear factors influencing crash occurrences through a novel Geospatial Explainable Artificial Intelligence (GeoXAI) framework using AutoML and GeoShapley. AutoML is an automated machine learning approach that streamlines model selection, hyperparameter tuning, and training to identify high-performing predictive models with minimal manual intervention (He et al., 2021). GeoShapley is a game theory-based method that extends from SHAP and can measure both non-linearity and spatial effects underling a ML model (Li, 2024). This enables the estimation of location-specific feature importance which can offering geographically explicit insights into how each factors influencing on crash risk and why crash risk varies across space.

Overall, the contributions of this study are threefold:

i) A unified GeoXAI framework for traffic safety modeling that jointly uncovers spatial heterogeneity and non-linear relationships in crash occurrences, which provides a more nuanced and geographically aware interpretation than existing methods.

ii) Quantify the influence of socio-demographic and roadway characteristics on census tract level crash occurrences and how various factors interact to shape crash density patterns across space. Some valuable insights and suggestions are provided for transportation safety planning and policy development according to the findings.

iii) To validate the robustness of this approach, we also benchmark our results against the other two widely adopted approaches: traditional interpretable ML models (i.e., using ML and SHAP) and spatial statistical models (i.e., using MGWR). This comparison, as far as we know, is the first comparison of the three approaches on empirical dataset and examines the similarity and differences of the results that offer complementary view of crash analysis.

The remainder of this paper is structured as follows. Section 2 reviews the relevant literature on crash contributing factors and the evolution of nonlinear and spatial modeling approaches. Section 3 introduces the study area, data sources, and the variables used. The analytical framework of GeoXAI is detailed in Section 4. The results are presented in Section 5, followed by a discussion of the findings and their policy implications in Section 6, and the final section concludes the study.

# 2. Literature Review

## 2.1 Traffic Crash Contributing Factors

The social and physical environment in which travel occurs including demographic, economic, and land use characteristics, profoundly shapes crash risk. Crashes are not random events but rather the outcome of a complex interplay of multiple factors (Merlin et al., 2020; Xiao et al., 2024). Population density and urbanization increase exposure and potential conflicts among vehicles, pedestrians, and cyclists (Qiao et al., 2020). Vulnerable age groups, such as children and older adults, face higher risks due to physical and cognitive limitations (Tournier et al., 2016; Wilmut & Purcell, 2022). Socioeconomic conditions, including income, poverty, and unemployment, often serve as proxies for underlying safety risks, such as poorly maintained vehicles, elevated driver stress, or residence in areas with inadequate infrastructure (Traynor, 2009; Sagar et al., 2021; Sohaee & Bohluli, 2024). Commuting patterns, including longer travel times and higher numbers of commuters, further increase exposure and fatigue, heightening crash likelihood (Elias et al., 2010; Cui et al., 2025).

Beyond demographics, the physical form of the built environment plays a decisive role. Traffic volume, typically measured by Annual Average Daily Traffic (AADT), influences crash frequency, often in a nonlinear manner, with both very low and very high traffic conditions associated with elevated crash rates (Abdel-Aty & Radwan, 2000; Wang et al., 2013). Roadway design elements, such as lane width, median width, intersection density, and the overall density of the road network, shape opportunities for vehicle interactions and affect crash occurrence (Papadimitriou et al., 2019; Xie et al., 2014; Lee et al., 2017). Traffic composition, especially the proportion of heavy-duty trucks, impacts both the likelihood and severity of collisions (Lemp et al., 2011). Speed management and enforcement infrastructure, including posted speed limits and speed cameras, directly influence driving behavior and crash severity, with higher speeds increasing crash energy and enforcement density promoting safer driving (Doecke et al., 2018; Shi & Abdel-Aty, 2015; Kabir et al., 2021).

Despite these extensive findings, many existing studies have focused primarily on examining how influencing factors affect the crash related dependent variables, such as crash frequency, while overlooking the nonlinear relationships and spatial effects these factors may have on crash density. This limitation highlights a critical need for innovative approaches that can capture these complex interactions. By addressing both nonlinearity and spatial heterogeneity, researchers can gain more precise understanding of how various factors contribute to crash occurrences, ultimately enabling the development of more targeted and effective safety strategies.

## 2.2 Nonlinear Relationships and Spatial Effects Modeling in Crash Modeling

Crash occurrence is a complex process, with relationships between influencing factors and crash rate often being nonlinear and spatially varying (Zeng et al., 2016; Azimian et al., 2021; Zhou et al., 2025). Therefore, accurately modeling these nonlinear patterns and spatial effects is essential for deeper insights into the underlying causes of crashes.

Traditional methods such as Poisson regression and Bayesian model serve as classic statistics tools for crash rate analysis. However, these methods share a key core limitation: they rely on the linear relationship assumption between explanatory variables and crash rates (Lord & Mannering, 2010; Savolainen et al., 2011). As a result, they are unable to effectively capture complex nonlinear patterns in crash data. Compared to traditional statistics approaches, the integration of ML method and SHAP allows for more comprehensive model non-linear relationship interpretation (Islam & Abdel-Aty, 2024). Recent studies have increasingly integrated ML models and SHAP to better uncover the nonlinear and complex mechanisms underlying crash occurrences. However, these non-linear relationship capturing models assume that the influence of variables remains constant across the entire spatial domain, which can obscure important local variations and overlook spatial effects (Shariat-Mohaymany et al., 2015).

To address the spatial heterogeneity effects of crash analysis, GWR have been widely used by fitting a separate local regression model at each location (Wang et al., 2019). Recently, Multiscale Geographically Weighted Regression (MGWR) was proposed to further improve the modeling of spatial heterogeneity. Unlike traditional GWR, which assumes that all variables operate at the same spatial scale, MGWR allows each predictor to vary across space at its own scale (Fotheringham et al., 2017). In crash analysis, MGWR consistently outperforms OLS and GWR by more effectively capturing the spatial heterogeneity of demographic and roadway variables associated with fatal/injury and property-damage-only crashes (Liu et al., 2024). With the potential for revealing spatial heterogeneity that remains hidden under global modeling approaches, Li et al. (2022) found that strong correlations between social vulnerability and crash outcomes in urbanized and border regions, with minority status and socioeconomic conditions emerging as critical spatial determinants. MGWR provided a substantially better model fit than traditional GWR, particularly by capturing factor-specific scales such as node counts, POI entrance densities, and terrain effects, which influenced crashes differently across urban subdistricts (Tang et al., 2023).

To combine the strengths of nonlinear modeling with spatial heterogeneity analysis, Li (2024) has proposed GeoShapley, a geospatial extension of SHAP. GeoShapley enables the quantification of the importance of location, as well as the interactions between location and other features within a model (Li, 2024). Its primary strength is its ability to treat geographic location as a unified feature and quantify not only its direct effect but also its interaction with every other variable in the model. By allowing these spatial interactions, GeoShapley simultaneously disentangles nonlinear relationships and spatial variability, offering a level of interpretability that traditional models cannot achieve. Applications in multiple domains have validated GeoShapley's effectiveness in understanding of the nonlinear relationship and spatial heterogeneity. For example, Peng et al. (2025) use it to uncover how truck-related $CO_2$ emissions are shaped by spatial infrastructure disparities and the nonlinear relationship between them. Chen et al. (2025) analyze nonlinear and spatially heterogeneous drivers of housing prices in Beijing. With GeoShaply model, Neto et al. (2025) examine how transport accessibility shapes retail location patterns by integrating real travel-time data for private and public transport. These studies demonstrate how GeoShapley provides robust, location-specific evidence that is essential for developing targeted, efficient, and equitable policies through decoding nonlinear relationships and spatial heterogeneity simultaneously.

# 3. Study Area and Data

## *3.1 Study Area*

This study focuses on the state of Florida, examining motor vehicle crash patterns that occurred from January 1, 2023, to December 31, 2024. As the third most populous state in the U.S., Florida features a diverse range of environments, including dense urban centers such as Miami and Tampa, expansive suburban communities, and rural regions with limited infrastructure. This diversity creates significant spatial heterogeneity in travel behavior, roadway design, and exposure to crash risk. The basic spatial unit of analysis employed in this study is the census tract, a standard geographic boundary defined by the U.S. Census Bureau (Ratcliffe et al., 2016). Census tracts offer fine spatial resolution and are consistently defined over time, making them well-suited for detailed crash risk assessment and integration with socio-economic and built environment data.

To construct a comprehensive analytical dataset, three main categories of data were collected and integrated for all census tracts across Florida: (1) traffic crash records, (2) roadway network and traffic characteristics, and (3) socio-demographic and economic factors.

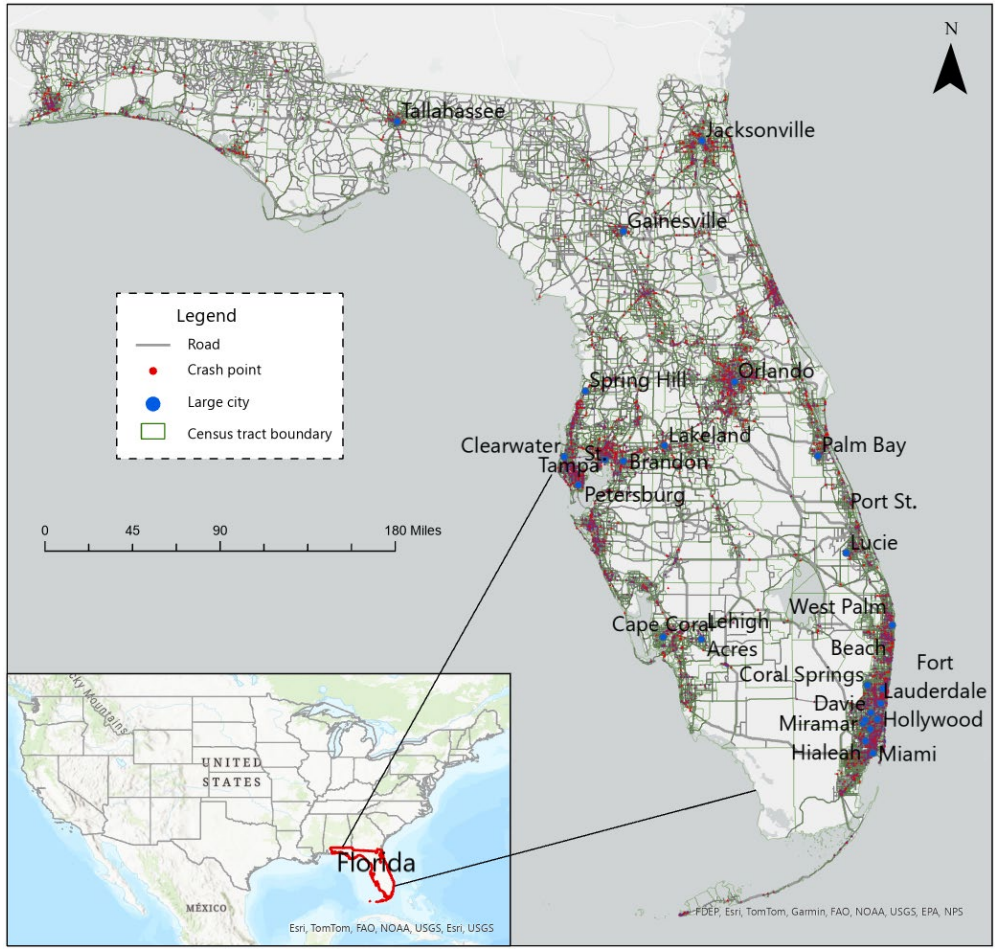

Figure 1: Crash occurrences in Florida

*3.2 Data Collection and Processing*

Crash records were obtained from the Signal Four Analytics (S4A) system maintained by the Florida Department of Transportation (FDOT). The S4A database is widely used in safety research and includes detailed information on crash severity, location, time, environmental conditions, and contributing factors, along with roadway and situational details. Crash density, the dependent variable in this study, is derived by spatially joining and aggregating the raw data at the census tract level. It is defined as the total number of crashes per square kilometer within each census tract.

Detailed information on roadway infrastructure and traffic characteristics is sourced from the Florida Department of Transportation GIS database (GIS-FDOT). This database is a comprehensive inventory containing information on all public roads in the state, including their location, functional classification, physical attributes, and traffic volumes. From this source, we derived key variables such as the density of different road classes (expressway, arterial, collector, and local), intersection density, traffic signal density, AADT for general traffic, average speed limits, road median width, road width, and the density of traffic monitoring sites. The road classification data is exacted and the length of different classes of roads counted in each census tract, after which we calculated the road density of different census tracts.

A wide range of socio-demographic and economic variables for each census tract is obtained from the U.S. Census Bureau's American Community Survey (ACS 2018-2023) 5-Year Estimates. The selected variables include population density, percentage of White population, percentage of unemployment, percentage of people with bachelor's degree or higher, and percentage of households without car. The compact neighborhood score is exacted from Housing and Transportation Affordability Index (H+T index), which provides a comprehensive view of affordability that includes both the cost of housing and the cost of transportation at the neighborhood level. Descriptions of all variables are presented in Table 2.

Table 2. List of Variables

| Variable | Unit | Description | Mean | St. Dev. | Data Source |
|---|---|---|---|---|---|
| **Dependent variable** | | | | | |
| Crash density | count/km2/year | Total number of crashes per km2 in each census tract | 55.35 | 102.44 | Signal Four |
| **Socio-demographic & Economic Characteristics** | | | | | |
| Percentage of white population | % | The proportion of White people in each census tract | 62.04 | 23.71 | ACS(2018-2023) |
| Percentage of people with bachelor's degree | % | The proportion of people with bachelor's degree in each census tract | 20.30 | 9.65 | ACS(2018-2023) |
| Population density | person/km2 | The number of people divided by area in each census tract | 1563.37 | 2178.86 | ACS(2018-2023) |
| Percentage of unemployment | % | The proportion of unemployment people in each census tract | 2.81 | 2.25 | ACS(2018-2023) |
| Percentage of households without car | % | The proportion of households without car in each census tract | 6.27 | 6.78 | ACS(2018-2023) |
| Compact neighborhood score | / | Compact neighborhood score (0-10) | 5.40 | 2.42 | H+T index |
| **Traffic & Roadway Characteristics** | | | | | |
| Intersections density | number/km² | The number of total intersections divided by area in each census tract | 15.34 | 18.67 | GIS-FDOT |
| Expressway density | km/km2 | The length of expressway divided by area in each census tract | 160.93 | 483.46 | GIS-FDOT |
| Arterial road density | km/km2 | The length of the arterial road divided by area in each census tract | 853.35 | 930.89 | GIS-FDOT |
| Collector road density | km/km2 | The length of the collector road divided by area in each census tract | 863.56 | 924.58 | GIS-FDOT |
| Local road density | km/km2 | The length of the local road divided by area in each census tract | 217.93 | 658.41 | GIS-FDOT |
| Average speed limits | mph | Average speed limits in each census tract | 43.27 | 8.41 | GIS-FDOT |
| Average road width | ft | Average road width in each census tract | 23.79 | 4.93 | GIS-FDOT |

| Traffic monitoring density | count/km2 | The number of traffic monitoring divided by area in each census tract | 10.74 | 17.72 | GIS-FDOT |
| AADT | vehicles per day | Annual average daily traffic | 21876.13 | 15633.55 | GIS-FDOT |

## 4. GeoXAI Framework

To investigate the complex nonlinear relationships and spatial patterns of traffic crash density, this study employs a multi-stage analytical framework. This framework shown in Figure 2 is designed to first identify the most effective predictive model for our dataset and then to deconstruct this model's predictions to yield interpretable, spatially aware insights. The process consists of two primary stages: (1) optimal ML model selection using an AutoML approach, and (2) in-depth model interpretation using the GeoShapley to explain nonlinear relationship and spatial heterogeneity.

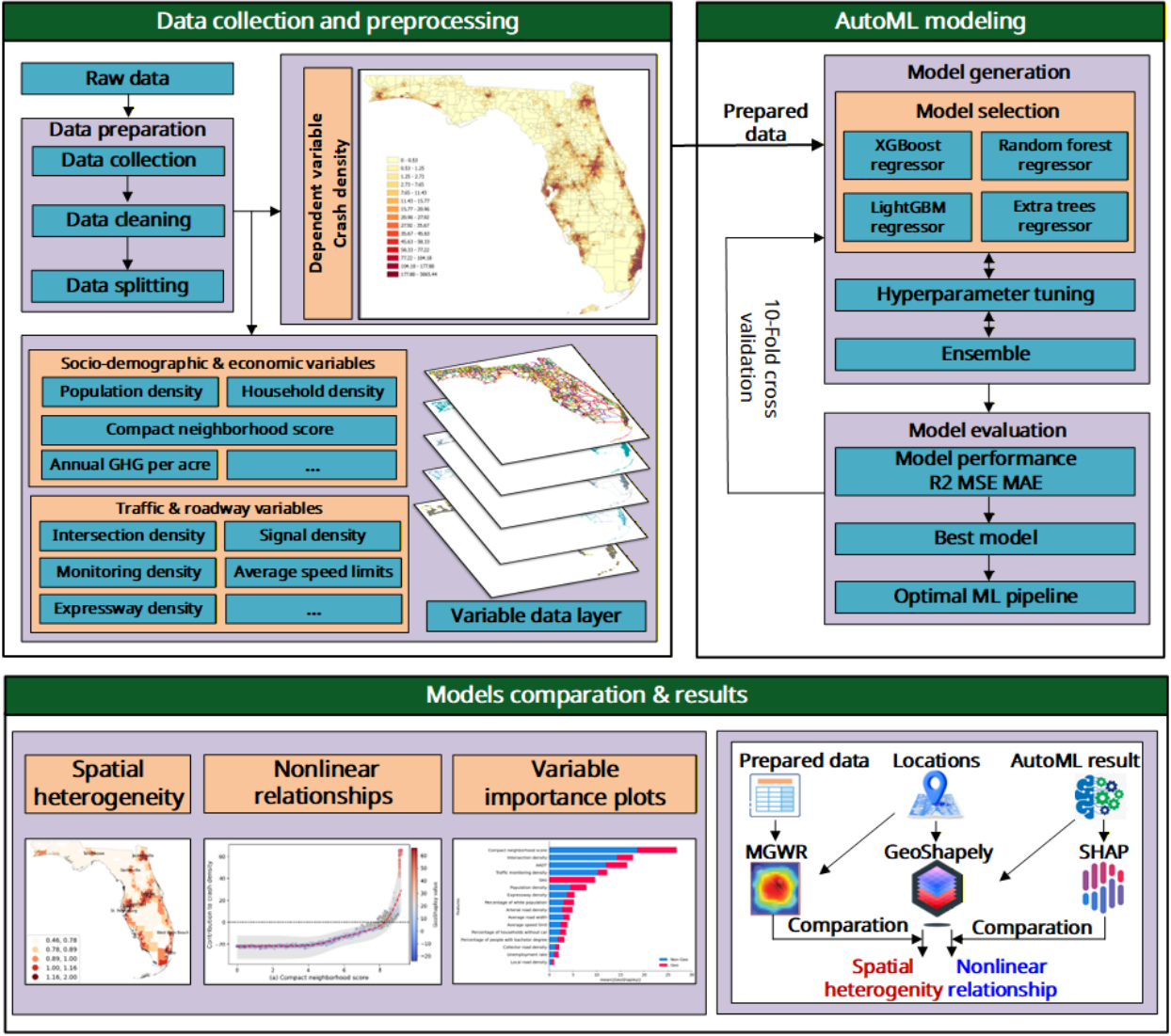

Figure 2. GeoXAI model framework

*4.1 AutoML modeling*

Given the multitude of ML algorithms available, selecting the most appropriate model can be a challenging and subjective process. To ensure strong predictive performance, this study adopts an AutoML framework that automates model selection and hyperparameter tuning (He et al., 2021). We utilize Fast Library for Automated Machine Learning & Tuning (FLAML), a lightweight and efficient AutoML tool developed by Microsoft Research (Wang et al., 2021), to identify the best regression model for predicting crash density at the census tract level. The process involves training and evaluating a suite of regression models, including Random Forest, XGBoost, LightGBM, and Extra Trees. The performance of each model is rigorously assessed using 10-fold cross-validation to prevent overfitting and ensure the model's generalizability to all data.

Formally, the AutoML algorithm aims to identify the optimal model's hyperparameter $\mathcal{M}^*$ that minimizes the expected prediction error on a validation dataset. This optimization problem can be expressed as:

$$\mathcal{M}^* = \arg\min_{\mathcal{M} \in \mathcal{H}} \frac{1}{n} \sum_{i=1}^{n} \mathcal{L}(y_i, \mathcal{M}(x_i)), \tag{1}$$

where $\mathcal{H}$ represents the space of candidate models and hyperparameter combinations, $(x_i, y_i)$ denotes the input–output pairs in the validation dataset, and $\mathcal{L}(\cdot)$ is a loss function that quantifies prediction error. FLAML operates by employing a greedy search strategy with adaptive resource allocation, enabling it to evaluate a wide variety of models within a limited time budget.

To ensure a robust and comprehensive evaluation of model performance, this study employs three commonly used regression metrics: Mean Absolute Error (MAE), Mean Squared Error (MSE, and the coefficient of determination (R²) (Chicco et al., 2021; Tatachar, 2021). Each of these metrics captures different aspects of model prediction quality, and together they offer a well-rounded understanding of accuracy, consistency, and explanatory power. MAE measures the average size of the prediction errors, without caring whether the model overestimates or underestimates. It's easy to understand and not too sensitive to big errors. MSE gives more weight to large errors by squaring them, which helps when we want to strongly avoid big mistakes, though it can be affected by outliers. R² represents the proportion of variance in the dependent variable that is explained by the model. An R² value of 1 indicates perfect predictions, while an R² of 0 indicates that the model explains none of the variability in the target variable. It is useful for understanding how well the model captures the underlying data patterns. These metrics are defined as follows:

$$\text{MAE} = \frac{1}{n}\sum_{i=1}^{n}|y_i - \hat{y}_i|, \tag{2}$$

$$\text{MSE} = \frac{1}{n}\sum_{i=1}^{n}(y_i - \hat{y}_i)^2, \tag{3}$$

$$R^2 = 1 - \frac{\sum_{i=1}^{n}(y_i - \hat{y}_i)^2}{\sum_{i=1}^{n}(y_i - \bar{y})^2}, \tag{4}$$

where $\hat{y}_i$ denotes the predicted value for the observation and $\bar{y}$ is the mean of observed values. These metrics are calculated on the cross-validation folds, and the model achieving the lowest average MAE while maintaining high $R^2$ is selected for downstream model explanation analysis.

*4.2 GeoShapley Explanation*

After obtaining the optimal predictive ML model with AutoML, we employ GeoShapley to explain the model's behaviors. The foundation of GeoShapley is rooted in Shapley value, a concept from cooperative game theory that provides a method for fairly distributing a payout (a model's prediction) among a coalition of players (the input features) (Lundberg and Lee, 2017). The GeoShapley value for a non-spatial feature $X_j$, denoted as $\phi_j$, can be computed by following:

$$\phi_j = \sum_{S \subseteq M \setminus \{j\}} \frac{s!(p-s-g)!}{(p-g+1)!} \left( f(S \cup \{j\}) - f(S) \right), \tag{5}$$

where $p$ is the total number of features, $M$ is the set of all features, $S$ is a subset of features not including $X_j$, and $f(\cdot)$ is represents the model's prediction given a subset of features.

The above value does not consider the joint effect of location coordinates and their interaction to other features when explaining a model with location information. The GeoShapley Model addresses this by conceptualizing location as a single, joint player in the prediction game. In our study, the latitude and longitude of each census tract's centroid are grouped into a single feature denoted as GEO by GeoShapley. In essence, in addition to a constant ($\phi_0$) and primary linear/non-linear feature effect $\phi_j$ as seen in normal SHAP, GeoShapley isolates two spatial components from model prediction ($\hat{y}$) as shown in Eq (6): 1) the intrinsic location effect ($\phi_{GEO}$) that quantifies the influence of geographic context to the model prediction, independent of other measured variables; and 2) the synergistic effect $\phi(GEO, j)$ between location ($GEO$) and a non-spatial feature $j$, revealing how the influence of that feature changes across different locations. The GeoShapley estimates can be computed by following:

$$\hat{y} = \phi_0 + \phi_{GEO} + \sum_{j=1}^{p} \phi_j + \sum_{j=1}^{p} \phi(GEO, j), \tag{6}$$

Both $\phi_{GEO}$ and $\phi(GEO, j)$ are critical components for understanding spatial heterogeneity within the predictive model, and Li (2024) found that these two spatial components' gradients with respect to the features are analogous to the spatially varying coefficients (SVC) in an (M)GWR model. In addition, $\phi_j$ can capture the location-invariant global linearity/non-linearity in the model. Specifically, GeoShapley value for location features can be estimated by Eqs. (7) and (8), respectively.

$$\phi_{GEO} = \sum_{S \subseteq M \setminus \{GEO\}} \frac{s!(p-s-g)!}{(p-g+1)!} \left( f(S \cup \{GEO\}) - f(S) \right), \tag{7}$$

where $g$ denotes the dimension of location features, typically set to 2 to represent two dimensional coordinates. When $g$ equals 1, the above equation reduces to the classic Shapley value. The synergistic effect between location and other feature ($\phi_{(GEO,j)}$) is calculated as:

$$\phi_{(GEO,j)} = \sum_{S \subseteq M \setminus \{GEO, j\}} \frac{s!(p-s-g-1)!}{(p-g+1)!} \Delta_{\{GEO, j\}}, \tag{8}$$

where $\Delta_{\{GEO,j\}} = f(S \cup \{GEO, j\}) - f(S \cup \{GEO\}) - f(S \cup \{j\}) + f(S)$.

The mathematical details of the benchmarked methods including SHAP and MGWR can be found in Appendix A and B.

## 5. Result

### 5.1 Model Evaluation

Several ML models, including RF, XGBoost, LightGBM, and Extra Trees, are evaluated within AutoML to identify the most effective predictive approach for crash density. Among these, XGBoost emerging as the optimal model. Performance is assessed using MAE, MSE and $R^2$ shown in Table 3. summarizes the comparative performance with performance from the MGWR model as the benchmark. For context, the observed crash density values range from 0.03 to 1141.00 count/km2/year, with a mean of 55.87 and a standard deviation of 91.55. Relative to this distribution, XGBoost's predictive errors are moderate relative to the distribution of crash density, with the MAE of 12.28 ($\approx$21.9% of the mean) and the MSE of 2604.22 ($\approx$23.% of the variance). The test $R^2$ of 0.88 indicates that the model explains 88% of the variance in crash density, which indicates a strong fit to the observed data. By comparison, MGWR produced $R^2$ of 0.81. ML model outperforms MGWR in predictive accuracy.

Table 3. Model performance comparison.

|        | MAE   | MSE     | $R^2$ |
|--------|-------|---------|-------|
| AutoML | 12.28 | 1239.23 | 0.88  |
| MGWR   | 20.43 | 1943.10 | 0.81  |

### 5.2 GeoShapley Result

#### 5.2.1 Variable importance

Figure 3 compares summary plot of the traditional SHAP and GeoShapley in interpreting feature contributions to crash density across Florida census tracts. In Figure 3b, the SHAP summary plot shows that variables such as compact neighborhood score, intersection density, traffic monitoring density and population density have the strongest influence on crash density, with predominantly positive impacts. Other factors such as AADT, arterial road density and expressway density show moderate positive contributions with the increasing of these features' values. Some features like expressway density and demographic characteristics (e.g., percentage of white population) have relatively modest but still meaningful effects. However, traditional SHAP does not explain any potential spatial heterogeneity. In contrast, GeoShapley provides a more nuanced interpretation by incorporating the spatial context values into the model explanation. The inclusion of the GEO feature represents how location and its intrinsic context influences the model prediction. Therefore, numerous spatial interaction terms (e.g., population density × GEO and intersection density× GEO) appear through integrating various features and 'GEO'. From the GeoShapley summary plot shown in Figure 3a, several spatial interaction terms (such as compact neighborhood score × GEO and AADT × GEO) emerge among the top contributors, which reveals substantial spatial heterogeneity that is completely masked in the SHAP analysis. Beyond these dominant features, GeoShapley also reshapes the relative importance of several non-spatial features compared with their SHAP rankings. Once spatial interactions are incorporated, AADT moves upward and in importance in the GeoShapley ranking. In contrast, features such as population density show a noticeable decline in influence. The reduction influences of the percentage of white population and expressway density are shown in the plot of GeoShapley compared with SHAP, while features such as AADT and compact neighborhood score exhibit altered rankings due to the contribution of their spatial interaction terms. This comparison shows though SHAP and GeoShapley produce similar overall feature importance rankings, GeoShapley enhances interpretability by uncovering localized feature effects, which offers deeper insights into spatially dependent crash determinants that are masked in the typical SHAP analysis.

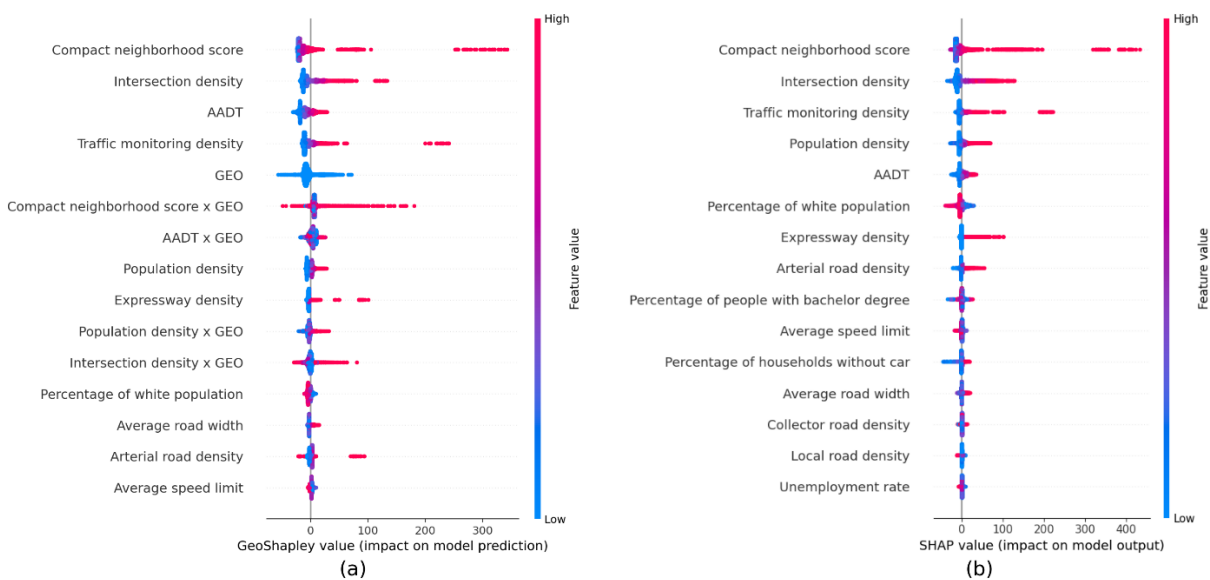

Figure 3. Comparation of GeoShapley and SHAP.

To understand global feature importance, Figure 4 provides a detailed breakdown of GeoShapley values, which decomposes each variable's influence from non-spatial (location-invariant effects) and spatial (location-specific effects) components. For each feature, the blue bar represents the average location-invariant effect, while the red bar captures the average location-varying effect. The accompanying donut charts summarize this division for the top 8 features, showing that the total contribution of location-invariant effects (80.93) is more than once that of location-specific effects (44.96). For the dominant location-invariant effects, the compact neighborhood score and intersection density are the most significant contributors. In the analysis of location-specific effects, the compact neighborhood score is again the leading factor, followed closely by the location-specific effects itself and the percentage of households without cars.

The compact neighborhood score emerges as the most powerful predictor by a significant margin. This feature not only has the largest overall impact but also exhibits the strongest interaction with geography, as shown by its substantial red component. This indicates that the effect of neighborhood compactness on the outcome is highly dependent on location and its spatial context. It is followed by intersection density, another strong predictor whose influence is primarily independent of geography. AADT ranks third, demonstrating a notable contribution from both its global effect and its spatial interaction. Traffic monitoring density and the intrinsic Geo variable round out the top five. Further down the ranking, population density and expressway density show moderate predictive power. In contrast, features such as unemployment rate and local road density have the least influence in this model.

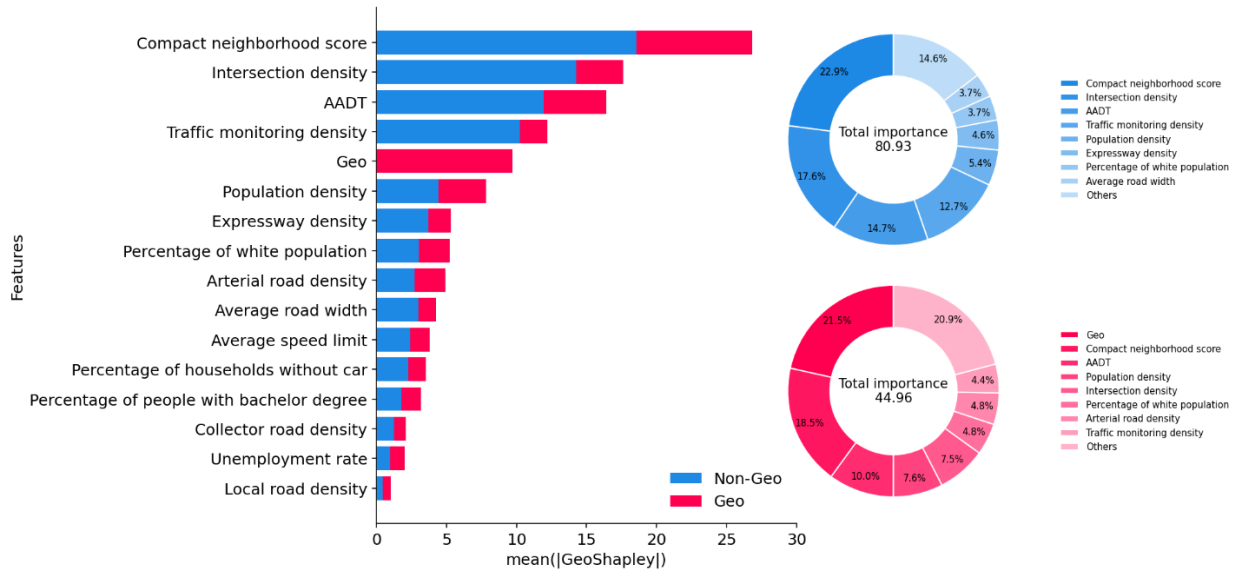

Figure 4. Global feature importance ranking.

### 5.2.2 Nonlinear relationship

The partial dependence of explanatory variables on crash density is shown in Figure 5. Each subplot illustrates how predicted crash density changes with the value of a given variable, while all other variables are held constant. To quantify the amount of model uncertainty in calculating the GeoShapley values, we conducted a bootstrap procedure following Li (2025), so to measure the individual 95% confidence intervals of the GeoShapley values. The overall confidence band is shown as grey, individual insignificant values are masked in white. Significant values are color-coded from blue to red according to magnitude.

The compact neighborhood score exhibits a sharp increase in crash density when the score exceeds 7, which shows that hyper-dense urban environments are strongly associated with higher crash risk. In contrast, crash risk is lowest in low- to moderately compact areas but escalates rapidly in dense urban cores with intense multimodal conflicts. Population density and the percentage of households without cars both demonstrate strong positive relationships with crash density. Greater population density increases exposure among all road users, naturally raising crash probability. Likewise, areas with fewer private vehicles tend to experience higher crash rates, which are often low-income areas. Several features display U-shaped patterns, indicating thresholds beyond which crash density changes direction. For instance, the percentage of the population with a bachelor's degree is associated with declining crash density up to around 35. However, crash density rises again at very high levels, which reflects the presence of universities and educational institutions concentrated in dense urban areas. The unemployment rate also shows a generally positive relationship with crash density. Interestingly, crashes are somewhat more frequent in areas with very low unemployment rates, which are often typically urban downtown areas. However, crash density increases further in high-unemployment areas, consistent with the expectation that these regions are often low-income. By contrast, the percentage of white residents exhibits a consistently strong negative association with crash density, with predominantly white census tracts experiencing substantially fewer crashes.

For traffic and roadway characteristics shown in Figure 5g-5o, variables such as AADT, intersection density, average road width, and traffic monitoring density show strong positive correlations with crash density. This pattern is consistent with greater vehicle exposure, more conflict points, and complex traffic control systems that increase the likelihood of crashes (Huang et al., 2010; Li et al., 2013). Regarding speed limits, crash density peaks around 35-45 mph, while higher limits are associated with fewer crashes, suggesting that certain highways may experience lower crash rates. From Figure 5i–5o, the influence of different road hierarchy levels on crash occurrences varies markedly. As the classification shifts from higher-order to lower-order roads, the positive association with crash density progressively weakens. This pattern continues until, at very high densities, local roads exert only a minimal or negligible impact on crash occurrences.

To validate the non-linear effects generated by GeoShapley, we also created partial dependence plots from the SHAP results, which are shown in Figure A1 (Appendix A). The non-linearities captured by SHAP are generally consistent with those explained by GeoShapley, but their shapes are mostly more pronounced in GeoShapley. This is because GeoShapley isolates spatial effects from non-linear effects, whereas SHAP combines all effects and attributes them to each feature, making any non-linearity less distinct.

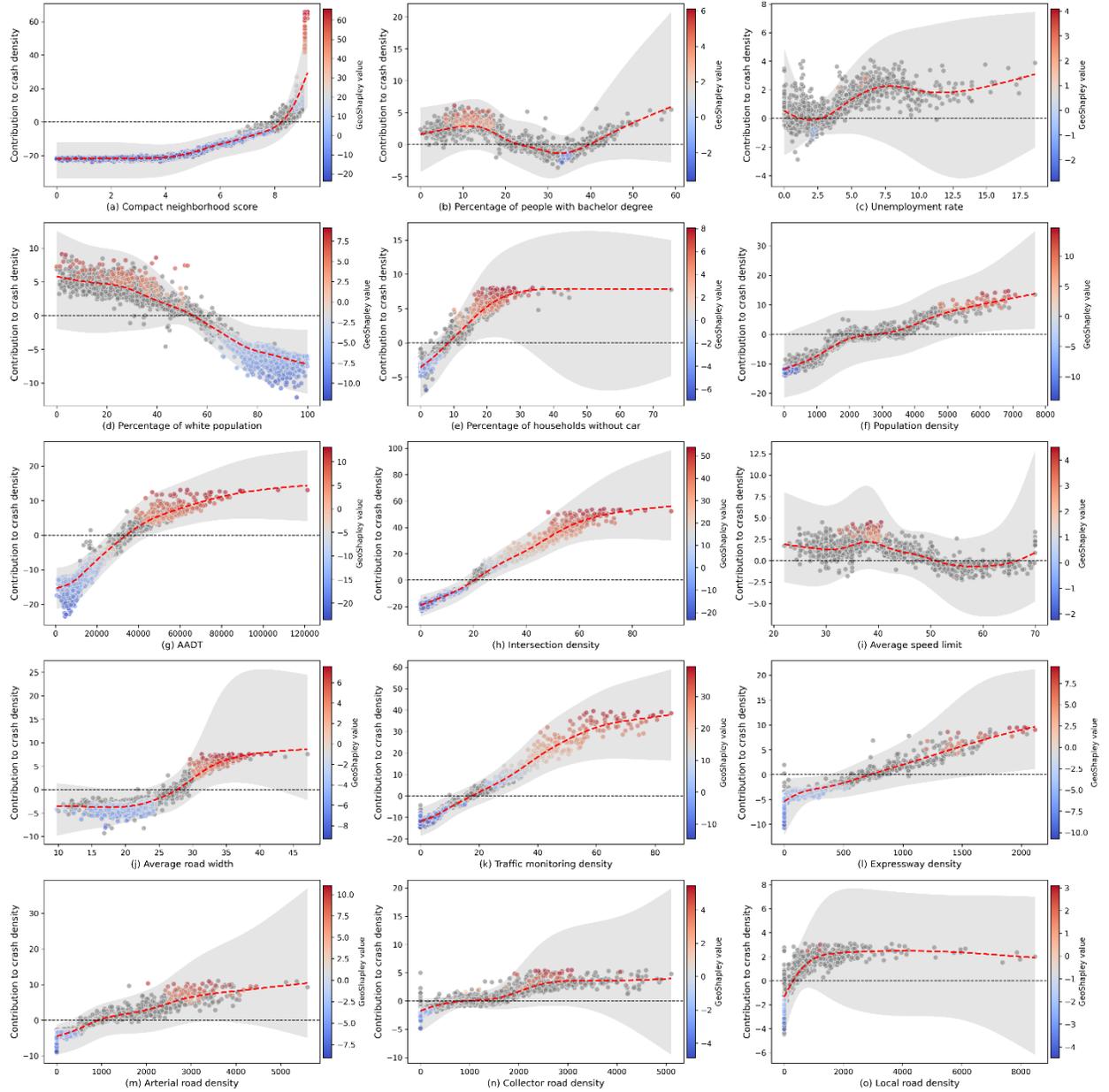

Figure 5. Nonlinear relationship between input variables and their contributions to crash.

### 5.2.3 Intrinsic location effect

Next we examine the spatial effects. As shown in Figure 6, the spatial distribution of the intrinsic location effect is illustrated using the SVC values from Eq. (7). The intercept represents the intrinsic contextual contribution of location when other features are held constant. A higher intercept indicates that the baseline crash density is inherently greater in urban areas than in rural settings. With GeoShapley shown in Figure 6a, the results reveal substantial heterogeneity across Florida census tracts. Northern and rural areas generally exhibit lower interceptive values, while urban areas, such as those near Miami, display higher baseline levels. Strong location effects are concentrated in South Florida, particularly around Miami and West Palm Beach, with additional clusters of relatively strong effects in Jacksonville, Tampa Bay and Orlando, and Tallahassee. Although urban areas tend to exert a positive influence on crash occurrence, the results also reveal notable intra-urban variation, particularly in the Orlando and Tampa Bay regions, suggesting that contextual crash risk factors are not uniform even within cities.

To compare and validate the contextual effects measured by GeoShapley, we also fit an MGWR model and mapped its local intercept, which represents the intrinsic contextual effect; this is shown in Figure 6b. As can be seen, the intrinsic contextual effects measured by MGWR exhibit broader regional patterns, but their magnitudes are generally weaker than those captured by GeoShapley. The MGWR results indicate elevated crash risk in Tallahassee, the Jacksonville metropolitan area, and Southwest Florida when other factors are controlled. Overall, both methods reveal consistent spatial patterns, but MGWR shows broader regional trends, whereas GeoShapley uncovers sharper and more localized contrasts even within each metropolitan area. This difference arises because MGWR employs a data-borrowing mechanism during local model fitting, which may oversmooth and average out fine-scale urban–rural differences, whereas GeoShapley and its underlying machine-learning framework do not impose this assumption.

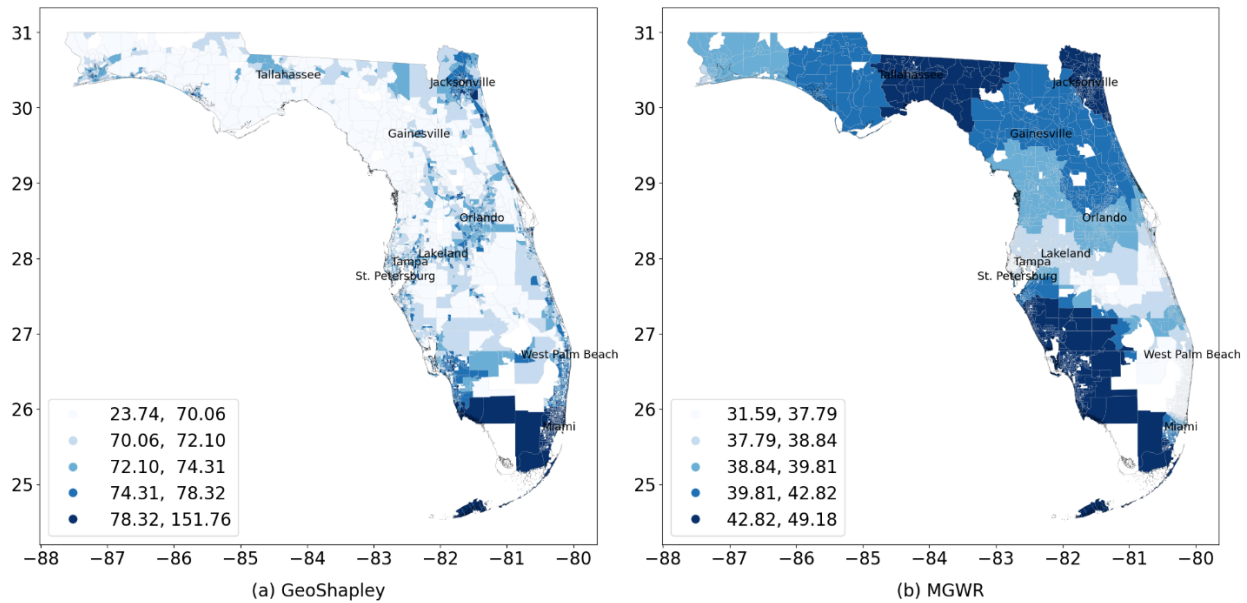

Figure 6. Intrinsic location effect estimated by GeoXAI model and MGWR.

### 5.2.4 Spatial heterogeneity

This section explores the geographic distribution of each factors marginal effect on crash density at the census tract level across Florida. Figures 7 and 8 illustrate how the influence of individual factors varies different census tracts.

(1) Socio-demographic & economic characteristics

The compact neighborhood score (Figure 7a) shows a much stronger positive effect in urbanized areas than in rural regions. Although compact urban design promotes walkability and sustainable land use, it can also elevate crash risks in dense city environments due to intensified multimodal interactions, narrower street configurations, and higher pedestrian activity. For the percentage of people with a bachelor's degree, a clear North–South divide emerges. In rural areas, higher educational attainment is strongly and positively associated with crash density. In contrast, the Southeast corridor (Miami and Fort Lauderdale) exhibits a strong negative relationship, suggesting that in these dense metro areas, higher education levels are linked to safer driving behavior and reduced crash risk. Unemployment rate shows scattered positive effects around major metropolitan regions, which indicates that economic disparities in urban–rural transition zones have a stronger influence on crash incidence than in the metropolitan cores. The percentage of white population displays a predominantly positive association with crash density across rural North and Central Florida. In sharp contrast, Greater Miami shows a pronounced negative relationship, reinforcing the distinctive demographic and traffic safety patterns of this region compared with the rest of the state. The percentage of households without a car exhibits stronger positive impacts in major metropolitan areas than in rural regions, suggesting that carless households face greater exposure and vulnerability to crash risks in large cities. Finally, population density (Figure 7f) demonstrates stronger positive effects in the northern, panhandle, and southern rural areas, rather than in Florida's dense urban centers. This pattern indicates that in less developed regions, increases in population density contribute more noticeably to crash risk, whereas in already saturated urban environments, additional density has a more neutral or less significant effect.

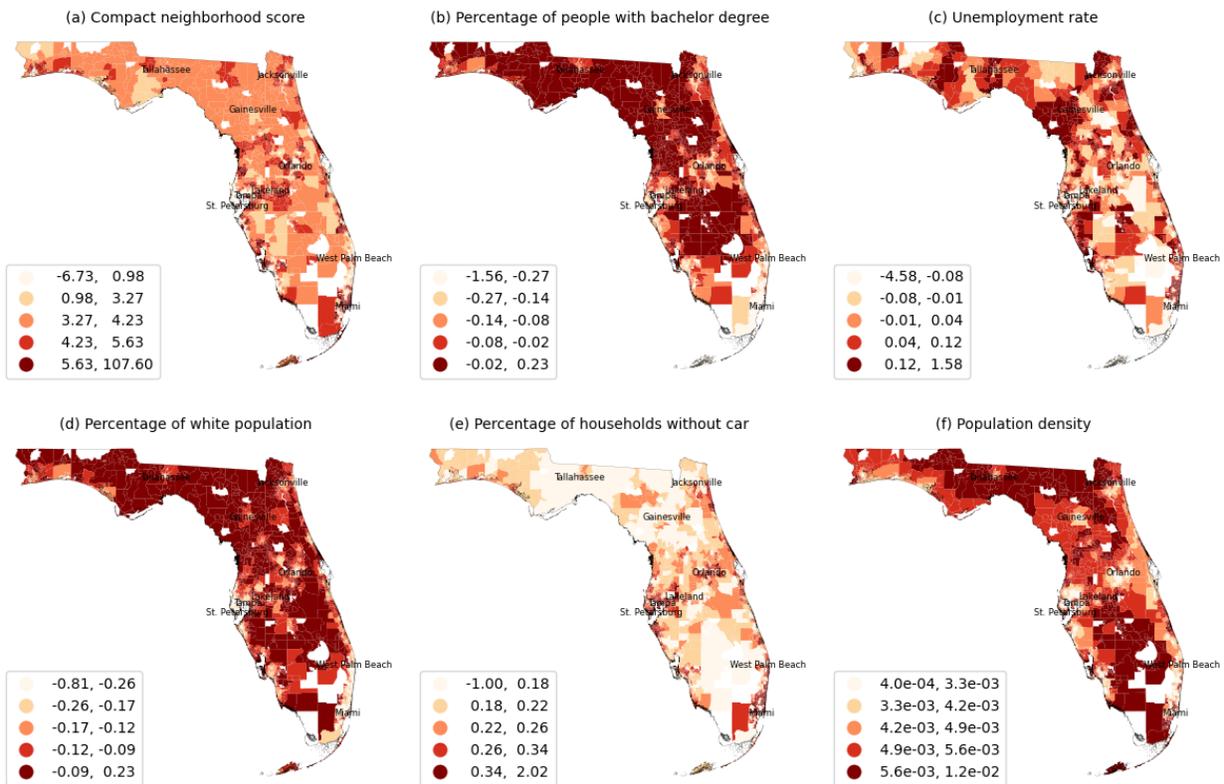

Figure 7. Spatially varying coefficients of socio-demographic & economic characteristics.

(2) Traffic & roadway characteristics

Several traffic and roadway variables exhibit strong spatial heterogeneity, particularly in major metropolitan regions such as Miami, Tampa Bay, and Orlando and the urbanized corridor extending from West Palm Beach to Miami.

AADT displays a widespread positive influence on crash density across the state. Stronger effects cluster around large metropolitan areas and the coastal urban corridor in Southeast Florida. This pattern reflects the elevated crash risk posed by consistently high traffic volumes along major regional travel corridors. Intersection density demonstrates strong regional variation. Highly urbanized areas such as Miami, Tampa, and Orlando exhibit notably higher intersection densities and correspondingly stronger positive effects on crash occurrences. Conversely, rural areas show weaker effects due to sparse intersection networks and lower conflict points. Higher average speed limits show a prominent positive effect primarily in Miami and surrounding dense neighborhoods. This is likely attributed to high-speed expressways like I-95 running through compact urban settings, where exposure to traffic conflict remains elevated despite higher design speeds. Broader roadways exhibit noticeable positive influences in major urban centers. Especially Miami and Orlando, areas with wider road experience increased crash incidence due to higher capacity roads accommodating substantial traffic flows. Traffic monitoring density reveals a strong spatial pattern tightly aligned with highly urbanized regions. The coastal corridor from Miami through West Palm Beach stands out as a hotspot, which indicates both a dense deployment of sensors and a concentration of crash-prone segments.

Expressway density shows the most limited spatial variation, appearing only in select metropolitan clusters, and its positive association with crashes remains relatively modest. As we move to arterial roads, density becomes more prominent in major urban centers such as Miami, Tampa, and Orlando. Collector roads exhibit a more diffuse suburban pattern, and their crash influence is weaker than that of arterials and expressways, reflecting their transitional role in channeling traffic between major corridors and local streets. Local roads display the most extensive and dense coverage across urbanized areas, particularly throughout Southeast Florida, and they show the weakest positive effect on crash density among all road types. Overall, as roadway classification shifts from higher-order to lower-order facilities, spatial density becomes more widespread while the associated crash effects gradually weaken.

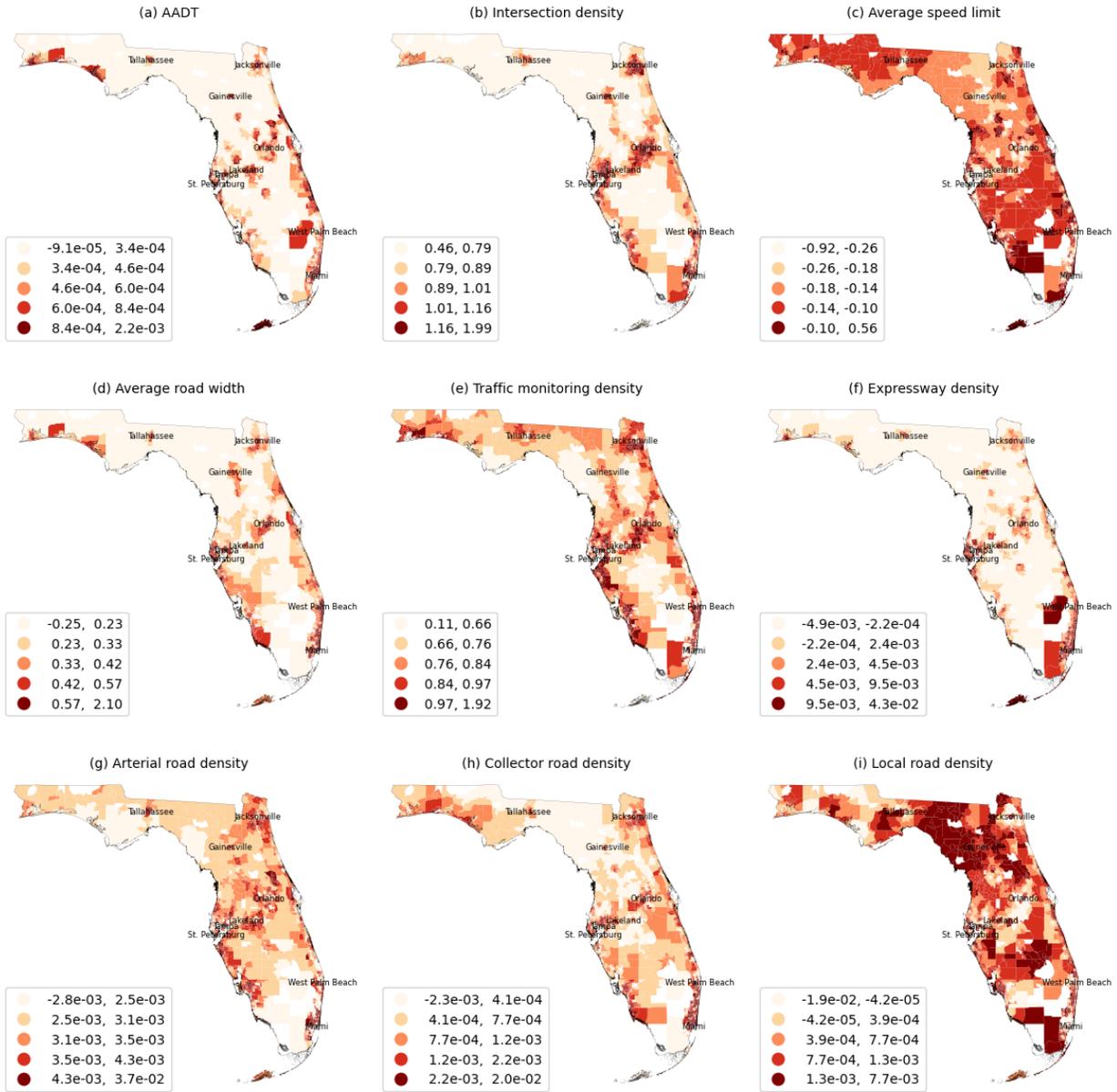

Figure 8. Spatially varying coefficients of traffic & roadway characteristics.

(3) Comparation with MGWR

We compare the spatial heterogeneity captured by GeoShapley and MGWR (shown in Appendix B) that exhibit local bandwidths in the MGWR results. MGWR captures spatial heterogeneity through explicitly varying local coefficients while GeoShapley reflects spatial heterogeneity through variations in local feature contributions derived from the result of AutoML. Overall, both methods reveal broadly consistent spatial trends for these features, indicating agreement in where stronger or weaker effects occur across the state. However, GeoShapley provides more detailed and nuanced spatial patterns, which reflects its ability to capture nonlinearities and interactions within the underlying ML model. As a result, while MGWR highlights the primary zones of spatial variation, GeoShapley uncovers finer-grained heterogeneity that enriches the overall spatial interpretation.

## 6. Discussion

This study employs a novel GeoXAI method to move beyond traditional crash analysis frameworks, which offers a deeper and more spatially nuanced understanding of the factors contributing to traffic crash density across Florida. By integrating a high-performance ML algorithm with a spatially explainable AI method (i.e., GeoShapley), the framework successfully deconstructs complex and nonlinear relationships and captures the spatial heterogeneity that characterizes traffic crash risk. The results not only enhance scientific understanding of traffic crashes but also provide actionable, location-specific insights for policymakers seeking to implement more effective and equitable safety interventions.

One of the most significant findings is the quantification of spatial heterogeneity in crash determinants. While conventional SHAP analysis treats variable effects as spatially uniform, the GeoShapley model decomposes each factor's impact into intrinsic and interaction-based components. It reveals that the influence of nearly every major factor varies considerably across Florida. Intrinsic contributions are pronounced in major metropolitan areas such as Miami, Orlando, Tampa, and Jacksonville, but relatively minor in rural regions. This implies that urban location itself imposes higher crash risk after controlling for local features. This pattern aligns with national statistics showing that 59% of motor vehicle traffic crashes in the United States occur in urban areas (Insurance Institute for Highway Safety, 2023), which indicates that urban locations account for most crashes. The large GeoShapley values in city tracts indicate that high traffic volumes, complex roadway networks, and dense land use inherently raise crash exposure. Conversely, minor intrinsic contributions in rural tracts reflect lower population density and traffic activity, which generally reduce the likelihood of crashes. These spatial interaction effects highlight that identical built environment factors, such as intersection density, can pose vastly different risks depending on geographic context. For instance, an increase in intersection density contributes far more to crash risk in downtown Orlando than the tracts in Florida's rural areas.

The study also revealed complex and nonlinear relationships in key built environment variables, particularly the compact neighborhood score. Moderate increases in neighborhood compactness are not associated with elevated crash risk, however, extremely high levels of compactness lead to a sharp rise in crash density. This pattern can be attributed to the intensified traffic flows and frequent pedestrians because of vehicle conflicts commonly observed in dense urban environments. The relationship is especially pronounced in central tracts of Miami-Dade County, where the compact urban fabric is strongly correlated with high GeoShapley contributions to crash predictions. These results are consistent with earlier findings that each 1% increase in metropolitan compactness reduces fatality rates by approximately 1.5% (Ewing et al., 2016), but also show that beyond a certain threshold, the safety benefits of compactness diminish or reverse. To mitigate these risks in dense urban areas, traffic calming and pedestrian-oriented design interventions should be prioritized. Effective measures include the implementation of neighborhood slow zones with speed humps, curb extensions, raised crosswalks, and enhanced signage. According to FHWA case studies, such zones have reduced vehicle speeds by 10–25% and crashes by approximately 10% (Federal Highway Administration Module 8). Additionally, installing pedestrian refuges and protected bike lanes can help safeguard vulnerable road users in compact districts.

Similarly, the density of local road and expressway displays a pronounced relationship with crash density. In areas with low roadway density, additional local roads have minimal effects. However, once expressway density surpasses a critical threshold, especially in high-speed, multi-lane environments, crash risk increases sharply. This finding is consistent with prior research linking high expressway presence with elevated crash rates (Dumbaugh et al., 2020). In our result, urban tracts with extensive road network showed significantly positive GeoShapley values. These results underscore the need for aggressive speed management in urban areas. One promising strategy is to gradually reduce posted speed limits in urban corridors and adjust rural limits upward where appropriate to better match the driving environment. This adaptive approach combined with automated speed enforcement and traffic calming, has demonstrated measurable reductions in crash frequency and severity (Khan & Das, 2024).

Another critical insight from this study concerns the location-dependent effects of intersection density and traffic monitoring. In high-density urban tracts, intersection density is one of the strongest predictors of crash risk due to the abundance of conflict points. By contrast, in rural areas where intersections are sparse and traffic volumes are low, intersection density has little to no impact on crash risk. This contrast is supported by national crash statistics: in 2022, only 16% of rural traffic fatalities occurred at intersections, compared to 32% in urban areas (Fatality Analysis Reporting System, 2024). While urban tracts with high levels of monitoring show elevated crash density because monitoring devices are concentrated in dense urban areas. Therefore, effective intersection management and enforcement require tailored solutions. In cities, planners should consider converting high-crash intersections to roundabouts, which the FHWA reports can reduce total crashes by 39% and injury crashes by 76% (Federal Highway Administration, 2006). Additionally, protected left-turn phases and pedestrian-only signal intervals can be used to reduce turning conflicts. In suburban and transitional zones, basic improvements such as improved sightlines, high-visibility crosswalks, and turn lanes should be prioritized. For rural crossroads with elevated crash rates, the addition of all-way stop control, advance warning flashers, or rumble strips can significantly improve safety. Equity-based enforcement such as increased patrols or targeted camera use in underserved neighborhoods also help reduce crash risk without resorting to punitive policies.

Socioeconomic variables, such as educational attainment and unemployment rate, are also found to have nonlinear effects. Crash density declines steeply as educational attainment increases from low to moderate levels but then plateaus at higher levels, reflecting diminishing safety returns beyond a certain point. This is consistent with prior findings that higher education levels correlate with safer driving behaviors (Sohaee & Bohluli, 2024). Similarly, the unemployment rate exhibits nonlinear effects, with moderate unemployment levels corresponding to higher crash rates, potentially due to reduced travel and exposure during severe economic downturns. These results underscore the need to address social vulnerability in transportation safety planning. Prior studies have documented the disproportionate crash burden borne by low-income and low-education communities (Dumbaugh et al., 2020). To mitigate these disparities, policymakers should fund targeted safety programs in these areas such as free child car seat distribution, multilingual safe-driving education campaigns, and investments in pedestrian infrastructure around schools. Engagement with community stakeholders through participatory planning can help ensure interventions are locally appropriate and effective.

## 7. Conclusion and Future Studies

This study introduces the novel GeoXAI framework, a methodological advancement that provides a more powerful and interpretable lens for traffic safety analysis. By simultaneously capturing the spatially heterogeneous and nonlinear determinants of crash density across Florida, the model overcomes the limitations of traditional approaches that often fail to account for these critical complexities. This research reveals nuanced insights into how built environment, socioeconomic, and roadway design factors contribute to crash occurrences in different geographic contexts. Some major findings and contributions are summarized below:

- Intersection density and traffic monitoring intensity are among the strongest contributors to crash risk in high-density urban regions.
- Nonlinear effects are observed in several key factors. For instance, crash density rises sharply in hyper-dense urban areas once neighborhood compactness exceeds a critical threshold, while unemployment rate exhibits a U-shaped pattern, decreasing crash risk up to moderate levels but increasing again in areas with very high educational concentrations.
- Socioeconomically disadvantaged communities experience disproportionately higher crash risk, which reveals an underlying spatial inequity in traffic safety outcomes.
- The pronounced spatial heterogeneity in crash risk underscores the importance of context-specific strategies, particularly in urban centers such as Miami-Dade, Orlando, Tampa, and Jacksonville.
- The GeoXAI framework enhances both interpretability and predictive accuracy, which offers a robust tool for identifying spatially varying crash determinants. The comparison and validation experiemenyts against alternative methods (SHAP and MGWR), shows that GeoShapley can exloan more nuanced and fine-grained spatial heterogeneity and clearer non-spatial non-linearity.

These findings suggest that in mitigating traffic crash risk, transportation and urban planning agencies should focus on the specific conditions contributing to elevated crash density in different geographic contexts. In dense urban areas, factors such as high intersection density, increased roadway compactness, and elevated arterial traffic volumes intensify crash risk. By identifying these contributing factors, agencies can implement targeted measures such as neighborhood traffic calming, intersection redesign, and adaptive speed enforcement to address the underlying causes of crashes in high-risk zones. Additionally, equity-based interventions, including community outreach and improved infrastructure in socioeconomically disadvantaged areas, can help mitigate the disproportionate burden of crashes on vulnerable populations. Furthermore, spatially informed planning can optimize the allocation of safety resources and reduce crashes more effectively. By tailoring interventions to the spatial patterns and nonlinear dynamics revealed in this study, policymakers can develop more responsive, efficient, and equitable traffic safety strategies.

It is worth noting that this study has several limitations that offer avenues for future research. First, the analysis is based on a static snapshot of crashes and built environment data; temporal variability (e.g., seasonal traffic patterns, weather factors) is not explicitly modeled. Incorporating temporal dynamics in future studies could improve understanding of how crash risk evolves over time. Second, our analysis is at the census tract level, some roadway characteristic variables, including roadway level, and speed limit, may lack sufficient spatial resolution or accuracy at this such scale. Future research could aim to incorporate a broader set of explanatory variables and to improve the granularity and reliability of roadway data in order to enhance model predictability and interpretability.

# Appendix A

As discussed in the GeoXAI framework, SHAP is used as a benchmarked method to compare with GeoShapley in capturing nonlinear relationships using the same AutoML results. In SHAP, each feature is treated as a "player" in a coalition game, and the model prediction is viewed as the game outcome (Shapley, 1953). The Shapley value represents the marginal contribution of a feature to the prediction, averaged over all possible feature subsets. Formally, the Shapley value for feature $X_j$ is:

$$\phi_j = \sum_{S \subseteq M \setminus \{j\}} \frac{s!(p-s-1)!}{p!} \left( f(S \cup \{j\}) - f(S) \right), \tag{A1}$$

where $p$ is the total number of features, $M$ is the set of all features, $S$ is a subset of all possible combinations of features not including $X_j$, $f(S)$ is the model output using features in $S$, and $f(S \cup \{j\})$ is the output with feature $X_j$ added. When applied to ML model, the "players" correspond to features, the model functions as the game, and the prediction is the outcome. Thus, Shapley values decompose an individual prediction into additive contributions from each feature (Štrumbelj and Kononenko, 2014), as expressed in Eq. (A2).

$$\hat{y}_i = \phi_0 + \sum_{j=1}^{p} \phi_{ji}, \tag{A2}$$

where $\phi_{ij}$ is the Shapley value of feature $X_j$ for observation $i$, and $\phi_0$ is the model's expected prediction.

Using SHAP, the nonlinear relationships between key explanatory variables and crash density are illustrated in Figure A1.

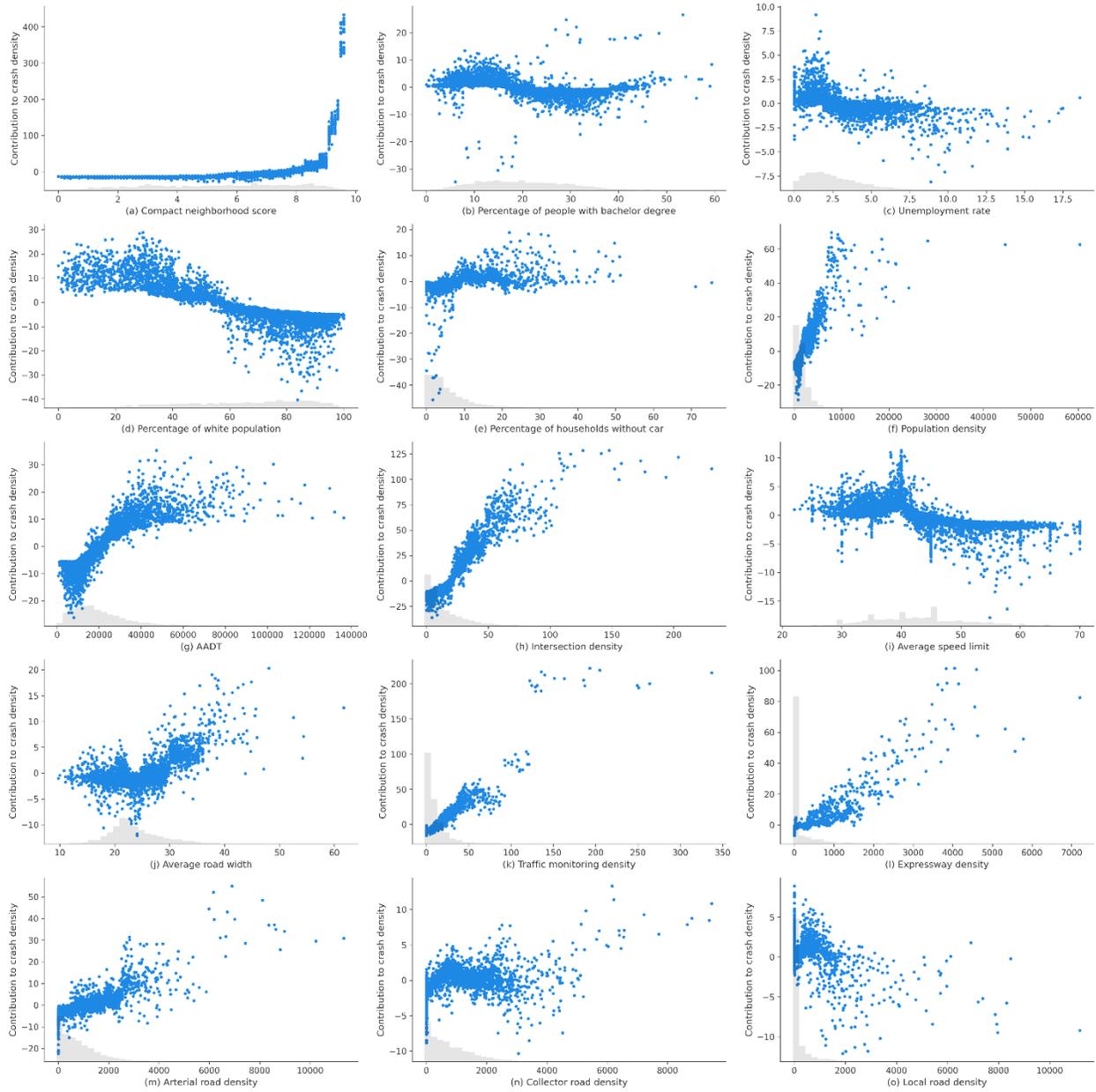

Figure A1. SHAP partial dependency plot.

# Appendix B

To provide a benchmark for evaluating spatial heterogeneity, an MGWR model is also estimated using the same dataset introduced in Figure 2. MGWR extends the traditional GWR framework by allowing each explanatory variable to vary at its own spatial scale, thereby capturing spatial processes that operate at different geographic extents. The analysis is implemented using the mgwr Python package (Oshan et al., 2019), which results are shown in Figure B1. As an established spatial modeling approach, MGWR offers a valuable point of comparison for GeoShapley (Li, 2025). While MGWR identifies spatial patterns through geographically varying coefficients, GeoShapley complements this by providing more detailed representation of spatial heterogeneity.

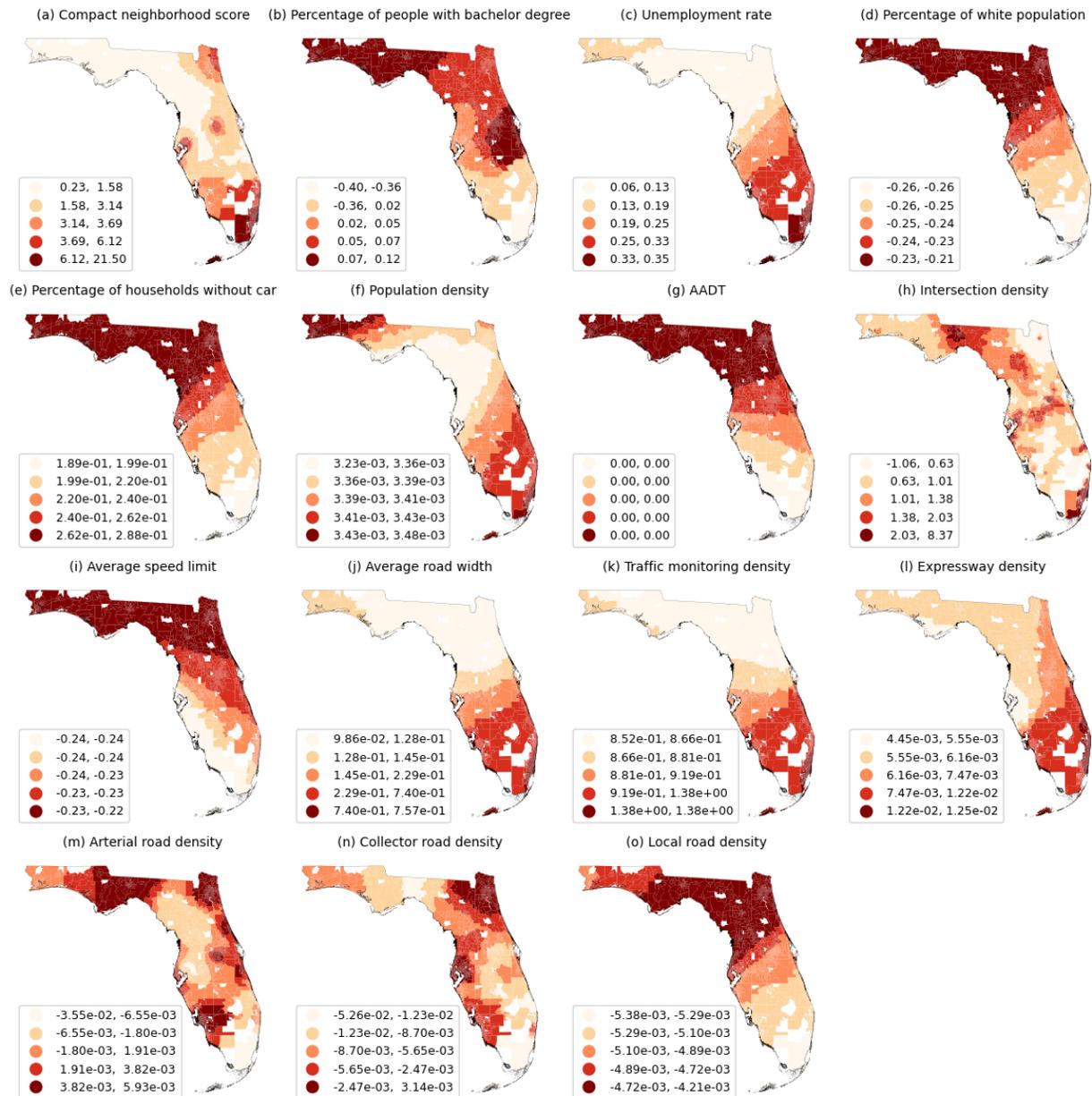

Figure B1. Spatially varying coefficients estimated from MGWR

Lemp, J. D., Kockelman, K. M., & Unnikrishnan, A. (2011). Analysis of large truck crash severity using heteroskedastic ordered probit models. *Accident Analysis & Prevention*, *43*(1), 370-380.

Li, X., Yu, S., Huang, X., Dadashova, B., Cui, W., & Zhang, Z. (2022). Do underserved and socially vulnerable communities observe more crashes? A spatial examination of social vulnerability and crash risks in Texas. *Accident Analysis & Prevention*, *173*, 106721.

Li, Z. (2022). Extracting spatial effects from machine learning model using local interpretation method: An example of SHAP and XGBoost. *Computers, Environment and Urban Systems*, *96*, 101845.

Li, Z. (2024). Geoshapley: A game theory approach to measuring spatial effects in machine learning models. *Annals of the American Association of Geographers*, *114*(7), 1365-1385.

Li, Z. (2025). Explainable AI in Spatial Analysis. In *GeoAI and Human Geography: The Dawn of a New Spatial Intelligence Era* (pp. 63-77). Cham: Springer Nature Switzerland.

Li, Z., Wang, W., Liu, P., Bigham, J. M., & Ragland, D. R. (2013). Using geographically weighted Poisson regression for county-level crash modeling in California. *Safety Science*, *58*, 89-97.

Liu, J., Khattak, A. J., & Wali, B. (2017). Do safety performance functions used for predicting crash frequency vary across space? Applying geographically weighted regressions to account for spatial heterogeneity. *Accident Analysis & Prevention*, *109*, 132-142.

Liu, J., Khattak, A. J., & Wali, B. (2017). Do safety performance functions used for predicting crash frequency vary across space? Applying geographically weighted regressions to account for spatial heterogeneity. *Accident Analysis & Prevention*, *109*, 132-142.

Liu, J., Das, S., & Khan, M. N. (2024). Decoding the impacts of contributory factors and addressing social disparities in crash frequency analysis. *Accident Analysis & Prevention*, *194*, 107375.

Lord, D., & Mannering, F. (2010). The statistical analysis of crash-frequency data: A review and assessment of methodological alternatives. *Transportation Research part A: Policy and Practice*, *44*(5), 291-305.

Lundberg, S. M., & Lee, S. I. (2017). A unified approach to interpreting model predictions. *Advances in Neural Information Processing Systems*, *30*.

Luo, Y., Yan, J., McClure, S. C., & Li, F. (2022). Socioeconomic and environmental factors of poverty in China using geographically weighted random forest regression model. *Environmental Science and Pollution Research*, 1-13.

Ma, Z., Zhang, H., Steven, I., Chien, J., Wang, J., & Dong, C. (2017). Predicting expressway crash frequency using a random effect negative binomial model: A case study in China. *Accident Analysis & Prevention*, *98*, 214-222.

Males, M. A. (2009). Poverty as a determinant of young drivers'' fatal crash risks. *Journal of safety Safety Research*, *40*(6), 443-448.

Man, C. K., Quddus, M., & Theofilatos, A. (2022). Transfer learning for spatio-temporal transferability of real-time crash prediction models. *Accident Analysis & Prevention*, *165*, 106511.

Merlin, L. A., Guerra, E., & Dumbaugh, E. (2020). Crash risk, crash exposure, and the built environment: A conceptual review. *Accident Analysis & Prevention*, *134*, 105244.

Neto, J. B. P., Santos, N. F., & Orrico Filho, R. D. (2025). Paths to prosperity: How transport networks and income accessibility shape retail location. *Journal of Transport Geography*, *128*, 104377.